\documentstyle[aps,epsf,preprint]{revtex}
\def\B.#1{{\bbox#1}}
\def\C.#1{{\cal  #1}}
\begin{document}
\title{Statistical conservation laws
in turbulent transport} \author{Itai Arad$^1$, Luca Biferale$^2$, Antonio
Celani$^3$, Itamar Procaccia$^1$ and Massimo Vergassola$^4$} 
\address{$^1$Dept. of Chemical Physics, The Weizmann Institute of Science,
Rehovot 76100, Israel. $^2$Dept. of Physics and INFM,
 University of Tor Vergata, Via 
della Ricerca Scientifica 1, I-00133 Roma, Italy. $^3$ CNRS, INLN, 
Route des Lucioles, 1361, 06560 Valbonne, France $^4$CNRS,
Observatoire de la C\^ote d'Azur, B.P. 4229, 06304 Nice Cedex 4, France}
\maketitle
\begin{abstract}
We address the statistical theory of fields that are transported by
a turbulent velocity field, both in forced and in unforced (decaying)
experiments. We propose that with
very few provisos on the transporting velocity field, correlation functions of the
transported field in the forced case are dominated by statistically preserved structures.
In decaying experiments (without forcing the transported fields)
we identify infinitely many statistical constants of the motion, which
are obtained by projecting the decaying correlation functions on the 
statistically preserved functions. We exemplify these ideas and provide
numerical evidence using a simple
model of turbulent transport. This example is chosen for its lack
of Lagrangian structure, to stress the generality of the ideas.
\end{abstract}

Turbulent transport represents a dynamical system in which
observable quantities (such as the concentration of a contaminant) change over time and
space in a  complex fashion \cite{SS00}. Unless the transported agent is continuously
renewed, irreversible mixing and dispersion result in a
uniform distribution over the system\cite{79MY,Csanady}. It had been therefore quite
surprising that in certain examples (in which the advecting velocity field is
$\delta$-correlated in time) it was  observed that there exists functions of space that
remain invariant to the time evolution \cite{CL,GK,ShSi}. In this Letter we propose that
with very few provisos on the turbulent velocity field,
the dynamics exhibits infinitely many statistically conserved
structures.  By ``statistically conserved" we mean structures that are
not conserved for every realization of the turbulent velocity field, but are
conserved after taking an average over all realizations of the turbulent
velocity field.  This
observation is very general, pertaining to scalar fields (like contaminants) or vector
fields (like magnetic fields), with pressure effects or without. Moreover, these
conserved structures dominate the statistics of the transported fields in the forced
case, (in which fresh supply is continuously furnished). Indeed, forced
experiments are more frequent in the turbulence community since they provide
much better statistical measurements than decaying experiments. The
present ideas, if supported by further studies, provide a
unifying approach to understanding such statistical properties in turbulent
transport, including the much debated issue of anomalous scaling \cite{Fri}.
\vskip 0.3 cm
Consider a turbulent velocity field $\B.u(\B.r,t)$ whose statistics is
assumed stationary, but without any further restrictions. We consider
the wide class of problems in which another field is transported
passively by that velocity. The field may be an advected scalar
$\theta(\B.r,t)$, with the equation of motion
\begin{equation}
\frac{\partial \theta}{\partial t} +\B.u\cdot \B.\nabla \theta=\kappa
\nabla^2\theta \ , \label{pas}
\end{equation}
or a vector, like a magnetic field $\B.B(\B.r,t)$ satisfying \cite{Zeldovich}
\begin{equation}
\frac{\partial \B.B}{\partial t} +(\B.u\cdot \B.\nabla) \B.B=(\B.B\cdot
\B.\nabla) \B.u + \kappa \nabla^2\B.B \ . \label{magnetic}
\end{equation}
We may also consider advection, as in \cite{01AP}, of a vector
field $\B.w$ whose divergence vanishes, $\nabla\cdot\B.w=0$:
\begin{equation}
\frac{\partial \B.w}{\partial t} +(\B.u\cdot \B.\nabla) \B.w=-\B.\nabla p
+ \kappa \nabla^2\B.w
\ .
\end{equation} 
Other examples can be
considered, as long as the equation for the transported field
$\phi(\B.r,t)$ (scalar or vector) has the form
\begin{equation}
\partial_t\phi={\cal L}\phi\, .
\label{basic}
\end{equation}
Here, ${\cal L}$ is a stochastic operator that is built out of the
turbulent velocity field. The key point is that the equation of motion
is {\em linear} in the transported field $\phi$. Statistical averages,
denoted as $\langle\dots\rangle$, are performed on the initial
conditions and on the realizations of $\B.u$. A fundamental
consequence of the linearity of the equations of motion is that the
correlation functions may be expressed as
\begin{equation}
\begin{array}{l}
\langle \phi(\B.r_1,t) \ldots \phi(\B.r_N,t) \rangle = \vspace{5pt}\\ 
\hspace{40pt} \int \C.P^{(N)}_{\underline{r},\underline{\rho}}(t)\, \langle
\phi(\B.\rho_1,0) \ldots \phi(\B.\rho_N,0) \rangle_0\, d\underline{\rho}
\ ,
\end{array}
\label{propagator}
\end{equation} 
where $\langle\dots\rangle_0$ on the RHS is an average over the statistics of 
the initial conditions. The notation
$\underline{\B.r}=({\B.r}_1,\ldots , {\B.r}_N)$ is used for simplicity. Note that we have
used the passive nature of the transported field, i.e. the fact that the velocity is
independent of the {\em initial} distribution of $\phi$, to separate the averages over the
initial conditions and the velocity. The linear operator
$\C.P^{(N)}_{\underline{r},\underline{\rho}}(t)$ propagates the $N$th-order correlation
function from time zero to time
$t$. \\

The evolution operator $\C.L$ generally includes dissipative terms, and without
fresh input (forcing) 
the statistics of the field $\phi$ is time-dependent; this is the {\em
decaying} case (\ref{basic}). A related problem of much experimental
and theoretical interest is {\em forced} turbulent transport where an
input term $f$ is added to the equation (\ref{basic}). The situations
of interest in turbulence typically involve an input acting only at
large scales of order $L$. The object of major interest is the
statistics of the transported field at distances much smaller than $L$ and at the
stationary state.  As usual in turbulent flows, the moments of the
velocity increments are assumed to behave as power laws with respect
to the distance between the two points.  The correlation functions are
then expected to contain anomalous contributions behaving as
\begin{equation}
\label{dilation}
\langle \phi(\lambda{\B.r}_1,t)\ldots\phi(\lambda{\B.r}_N,t) \rangle
=\lambda^{\zeta_N} \langle \phi({\B.r}_1,t)\ldots\phi({\B.r}_N,t)
\rangle ,
\end{equation}
with scaling exponents $\zeta_N$ which cannot be inferred from
dimensional analysis. \\
  
The aim of this Letter is to draw attention to two crucial ideas, without further
provisos on the velocity field. We do not prove these ideas here, we offer
them as conjectures, and will
exemplify them below using a simple example that strongly differs
from the known previous cases. Specifically, the model discussed below lacks the Lagrangian
structure heavily relied upon in previous works. This offers evidence for
the generality of the following ideas:
\vskip 0.2 cm
\noindent (i) In the decaying cases, despite the non-stationarity of
the statistics, there exist special functions $Z^{(N)}(\underline{\B.r})$
such that
\begin{equation} 
I^{(N)}(t) = \int Z^{(N)}(\underline{\B.r})\, \langle \phi(\B.r_1,t) \ldots
\phi(\B.r_N,t) \rangle \, d\underline{r} \;
\label{integrals}
\end{equation}
are statistical integrals of motion. In the limit of infinitely large
system it does not change
with time. In a finite system, and see Fig.1 as an example, it is constant
in time until some outer time scale $T_L$ is reached. It follows from
(\ref{propagator}) and the conservation of
$I^{(N)}(t)$ that in the infinite size limit the $Z^{(N)}$'s are statistically conserved
structures:
\begin{equation}
Z^{(N)}(\underline{\B.r})=\int
\C.P^{(N)}_{\underline{\B.\rho},\underline{\B.r}}(t)
Z^{(N)}(\underline{\B.\rho})\,d\underline{\B.\rho} \ . \label{eig}
\end{equation}
\vskip 0.2 cm
\noindent (ii) The anomalous part of the stationary correlation
functions in the forced problems is dominated by statistically
conserved structures.  A direct
consequence is that the small-scale statistics of the transported field $\phi$ in the
forced case rests on the understanding of the decaying problem. A by-product is that the
scaling exponents $\zeta_N$ are universal, i.e.  independent of the
forcing mechanisms for any forcing that is statistically independent of
the velocity field. 

To appreciate the generality of those statements it is desirable to
review briefly the development of the ideas leading to them. The first
instance \cite{CL,GK,ShSi} in which the role of statistically conserved structures
emerged was passive scalar advection by a velocity field with a short
correlation time and self-similar spatial correlations.  This is the
well known Kraichnan model \cite{RHK}, which is non-generic, but it has the
advantage that it lends itself to analytic calculations. In the
Kraichnan model the simultaneous $N$th order correlation function
satisfies a linear differential equation, which is inhomogeneous in
the presence of the forcing. The general solution is the sum of the
inhomogeneous and the homogeneous parts.  It turns out that both
exhibit scaling properties, but the scaling exponent of the
homogeneous part (the zero modes) is leading (smaller) compared to
the other.  Moreover, the subleading inhomogeneous exponent is
predictable by dimensional analysis \`a la Kolmogorov, whereas the
leading exponent is anomalous, and calls for an explicit calculation.
In the case of the Kraichnan model it is also clear that the
homogeneous part of the linear operator governs the rate of change of
the correlation function in the decaying case, and thus the zero modes
are statistically conserved structures for the decaying problem.

Next, the understanding of the nature of the conserved structures
followed from the analysis of the dynamics of {\em groups} of
Lagrangian trajectories of tracer particles \cite{98BGK,GZ,98FMV,98GPZ}.  The remark is
that for the passive scalar equation (\ref{pas}) the transported field is
conserved along the trajectories of the tracer particles
$d{\B.R}(t)=\B.u({\B.R}(t),t)\,dt+\sqrt{2\kappa}\,d\B.\beta(t)$, where
$\B.\beta(t)$ is a Brownian process. To know the scalar
field at position $\B.r$ and time $t$ it is enough to track the
corresponding tracer particle back to initial position $\B.\rho$.  The
evolution operator $\C.P^{(N)}_{\underline{r},\underline{\rho}}(t)$ in
(\ref{propagator}) coincides then with the probability density that
$N$ tracer particles reach the positions $\underline{\B.r}$ at time
$t$ given their initial positions $\underline{\B.\rho}$. For example,
to understand the exponent $\zeta_3$ one needs to focus on the
dynamics of three tracer particles.  Obviously, three particles define
at any moment of time a triangle, which in its turn is fully
characterized by one length scale $R$ (say the sum of the lengths of
its sides), two of its internal angles, and all the angles that
specify the orientation of the triangle in space.  When the particles
are advected by the turbulent velocity field, the scale $R$ of the
triangle and its shape (angles) change continuously. The statement
that can be made is that {\em there exist distributions on the space
of the triangle configurations, that are statistically invariant to
the turbulent dynamics} \cite{98BGK,GZ,00AP}. In other words, if we release trios of
Lagrangian tracers many times into the turbulent fluid, and we choose
the distribution of their shapes and sizes correctly, it will remain
invariant to the turbulent advection. Such statistically conserved
structures are the aforementioned zero modes and they come to dominate
the statistics of the scalar field at small scales. The anomalous
exponents of the zero modes, such as $\zeta_3$, can be understood as
the rescaling exponents characterizing precisely such special
distributions. Of course, the same ideas apply to any order
correlation function with the appropriate shape dynamics. The
relevance of Lagrangian trajectories can be also demonstrated for
the magnetic field case (\ref{magnetic}), by adding a tangent vector
to the tracer particle, and see \cite{CM} for more details.

All this was established for the non-generic Kraichnan model of
passive scalar advection. Generalizing the zero-mode ideas to generic 
cases has a clear importance, but it calls for coping with two different
problems. First, if the velocity field in not $\delta$-correlated in time, we have 
no time-independent operators anymore, and the idea of zero modes as
functions annihilated by a fixed operator cannot be carried over. The correlations functions
still obey linear equations of motion, but the non-vanishing correlation time
of the velocity field statistically couples different times. It is therefore
not obvious that single-time (simultaneous) objects might still be
statistically conserved. The first
indication in favor of the generality of the picture came in \cite{00CV}. 
The velocity field is
generated by the 2-dimensional Navier-Stokes equations (in the inverse cascade regime)
and has a finite correlation time. It is clear from the previous discussion that
breaking the non-generic approximation of $\delta$-correlation in time is the 
crucial point. The numerical evidence is that statistically conserved structures
are still present.
Two parallel things have been done: on the one hand, the third order correlation function
$\left<\phi(\B.r_1)
\phi(\B.r_2)
\phi(\B.r_3)\right>$ was measured in forced simulations. This
yielded the distribution expected to be invariant. On the other hand,
trios of Lagrangian trajectories were released into the flow, and it
was demonstrated that their evolution agreed with the notion of
statistical invariance.

The second problem is that the formulations proposed up to now heavily
exploit the Lagrangian structure of the dynamics. In this Letter we take
the ideas further and demonstrate that their generality transcends the 
applicability of Lagrangian
trajectories. The moment that we consider other models of transport, e.g.  with
pressure, the properties of $N$th order correlation functions can no
longer be connected to $N$ Lagrangian tracer particles. A new formulation
of the statistical conservation laws is thus needed. That is what (\ref{integrals})
provides for. It is clear from it that the key point is
not the Lagrangian structure of the correlation function propagation but that the linear
operator is associated with (statistically) preserved functions (cf. Eq.(\ref{eig})). The
preserved functions then define the integrals of motion. Of course, this still does not
necessarily mean that they also dominate the field correlations in the forced case. To
prove it in full generality is difficult, and certainly beyond the scope of this Letter.
Rather, we will demonstrate the point with a non-trivial example.

To address the issue in a model as different as possible from those 
considered so far, we analyze shell models, where the discretization of the phase
space destroys any Lagrangian structure. A possible discretization of the field
equation (\ref{basic}) reads \cite{Luca}
\begin{eqnarray}
{d\phi_m\over dt}&=&i\big(k_{m+1}\phi_{m+1}u_{m+1}
+k_m\phi_{m-1}u^*_{m}\big)-\kappa k_m^2\phi_m ,
\label{passive}\\
&\equiv&{\cal L}_{m,m'} \phi_{m'} \nonumber
\end{eqnarray}
where the $u_n$ variables are
generated by the ``Sabra'' shell model \cite{Sabra}
\begin{eqnarray} \label{sabra}
\frac{d u_n}{dt}&=&i\big( ak_{n+1}  u_{n+2}u_{n+1}^*
 + bk_n u_{n+1}u_{n-1}^*  \\ \nonumber
&& +ck_{n-1} u_{n-1}u_{n-2}\big)  -\nu k_n^2  u_n +f_n\,,
\end{eqnarray}
where the coefficients $a$, $b$, and $c$ are real. In our simulations
$\kappa=\nu=5\times 10^{-7}$, $a=1$, $b=-0.4$, and $c=a+b$.  
The wavevectors are $k_n=k_0\,2^m$ with $n=0,\dots,N$. The smallest wavevector
is given by $k_0=0.05$ while $N$ defines the ultraviolet cut-off.
 The velocity
forcing $f_n$ is limited to the first  shell $n=0$.
For $\kappa=\nu=0$ the energies $\sum_n |u_n|^2$ and $\sum_n |\phi_n|^2$ are
{\em dynamically conserved}, i.e. realization by realization. The operator $\C.P^{(N)}$
of Eq.(\ref{propagator}) takes here the explicit form 
$\langle R(t|t_0)\dots R(t|t_0)\rangle$
where $R(t|t_0)\equiv T^+\exp[\int_{t_0}^{t}ds {\cal L}(s)]$,
with $T^+$ being the time ordering operator.

To demonstrate the {\em statistical} conservation laws, two
things were done. First
we considered the forced problem, adding random forcing to the first  shell of
(\ref{passive}). We have measured all the available correlation functions of
second, fourth and sixth order. Due to phase symmetry constraints, these are
(we put a subscript $f$ to stress that these are statistical averages  in
the stationary forced ensemble):
\begin{eqnarray}
F_2(n)&\equiv& \langle |\phi_n|^2\rangle_f\ , \\
F^{(1)}_4(n,m)&\equiv&  \langle |\phi_n|^2|\phi_m|^2\rangle_f\ , \\
F^{(2)}_4(n)&\equiv&  \langle \phi_{n+2} \phi_{n+1}^* \phi_{n+1}^*
\phi_{n-1}\rangle_f\ , \\ F_6(n,m,k)&\equiv&  \langle |\phi_n|^2|\phi_m|^2
|\phi_k|^2\rangle_f \ .
\end{eqnarray}
Secondly, we studied the decaying problem,
 preparing initial states $\phi_n(t=0)$
and following their evolution. As initial states we took distributions of
$\phi_n=0$ except for $n= 14,15$ where we initialized the 
field with a constant modulus and random phases.
 We then computed the following objects:
\begin{eqnarray}
I^{(2)}(t)&\equiv& \sum_n \langle |\phi_n|^2\rangle(t) ~F_2(n)\ ,\\
I^{(4)}(t)&\equiv& \sum_{n,m} \langle |\phi_n|^2|\phi_m|^2\rangle(t) ~F^{(1)}_4(n,m)
\nonumber\\&+& \sum_n \langle \phi_{n+2}^* \phi_{n+1} \phi_{n+1}
\phi^*_{n-1} \rangle(t)  ~F^{(2)}_4(n)\ ,\\
I^{(6)}(t)&\equiv& \sum_{n,m,k} \langle |\phi_n|^2|\phi_m|^2|\phi_k|^2\rangle(t) ~F_6(n,m,k)
\end{eqnarray}

Fig.1 summarizes the results. We show, for these three orders, (i) the time dependence
of the $n$th order decaying correlation functions themselves, (ii) the time dependence
of $I^{(N)}(t)$. In panel (C) we show also for comparison the time dependence of
$I^{(6)}(t)$ if we replace the measured forced $F_6$ by its dimensional shell dependence
(i.e. the shell dependence if the Kolmogorov theory were right). We see that only the
properly computed
$I^{(n)}(t)$ are time independent for times smaller than the large
scale eddy turn over time,  $T_L$.
The decay observed for  times larger than $T_L$ is simply due to finite
size effects intervening when the decaying field reaches the largest
scales. Indeed, the correlation functions of the forced
problem (that we used for $Z^{(N)}$) are not dominated by the inertial
contributions at those scales. 

In conclusion, the importance of statistical conservation laws for turbulent
transport was demonstrated. We have exemplified both the existence of statistically
conserved objects {\em and} the fact that they come to dominate the small-scale
behavior of the forced stationary correlation functions. The only provisos are that
the turbulent velocity field has the usual scaling properties observed in turbulence,
and that the transported field is passive. That offers a unifying picture
to the statistical theory of turbulent transport. We propose that further
research, both numeric and analytic, can contribute to the solidification of the 
generality of these ideas.
\acknowledgements
This work had been supported in part by the European Commission under the
Research Training Network HPRN-CT-2000-00162. IP acknowledges partial support
by the German Israeli Foundation and the Naftali and Anna 
Backenroth-Bronicki Fund for Research in Chaos and Complexity. 
MV acknowledges useful discussions with G.~Falkovich and K.~Gaw\c{e}dzki.

%
%

\begin{figure}
\epsfxsize=7 truecm
\epsfbox{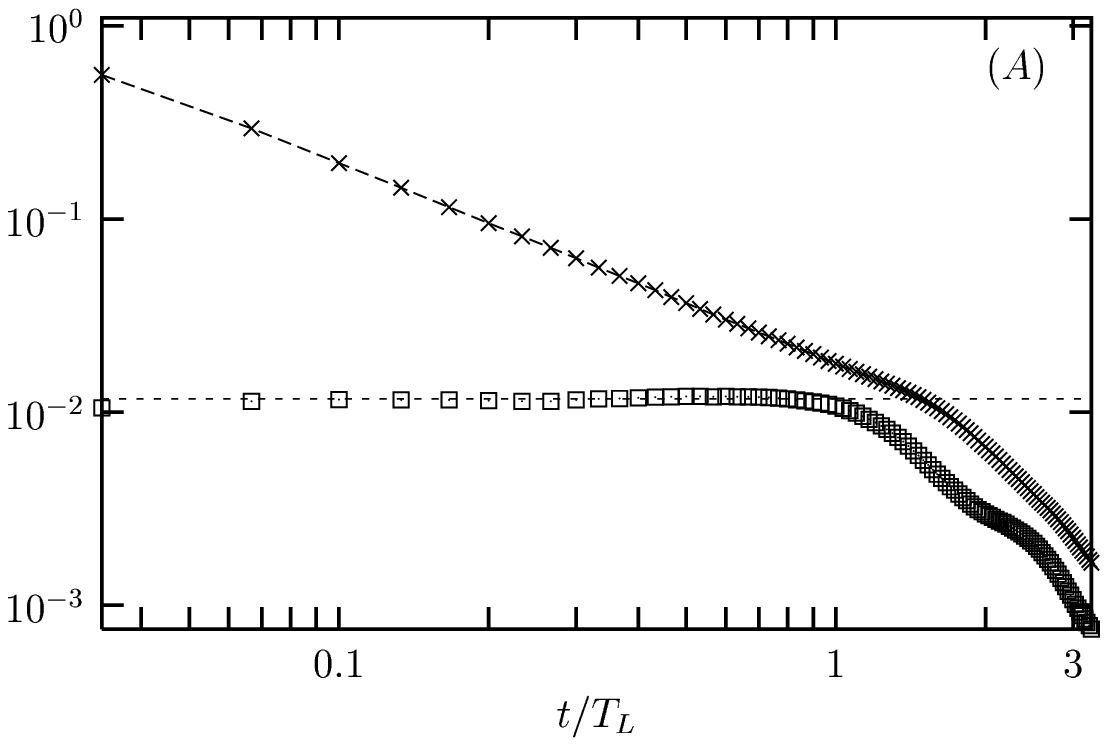}
\epsfxsize=7  truecm
\epsfbox{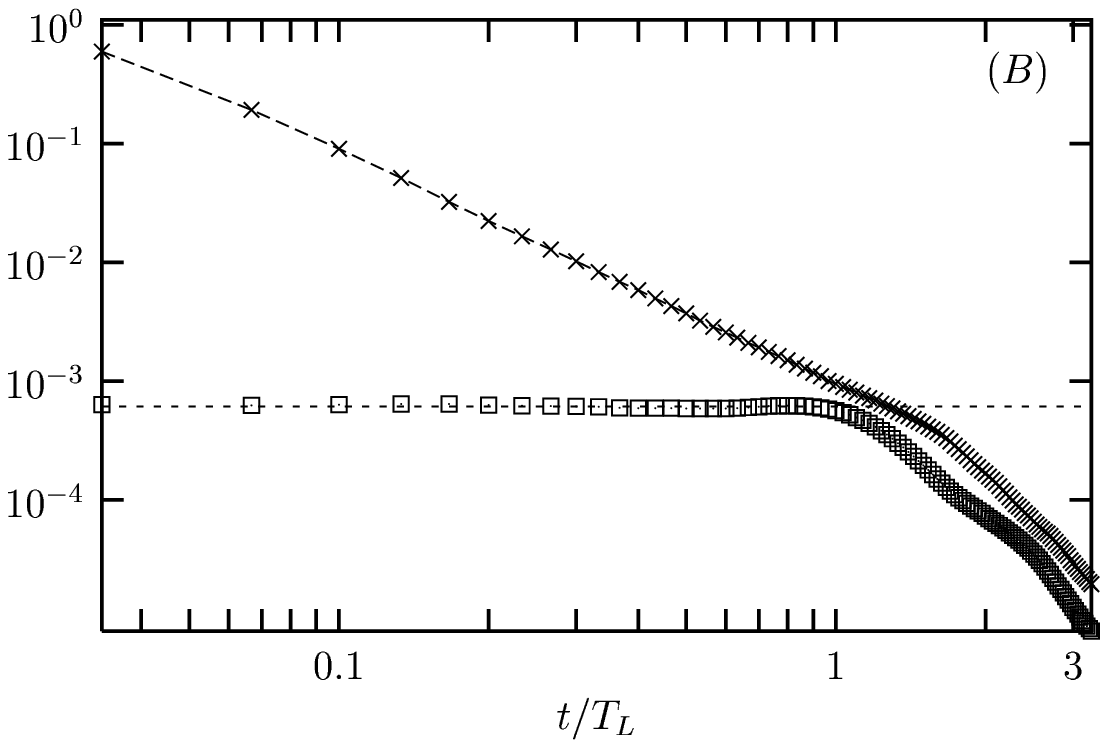}
\epsfxsize=7 truecm
\epsfbox{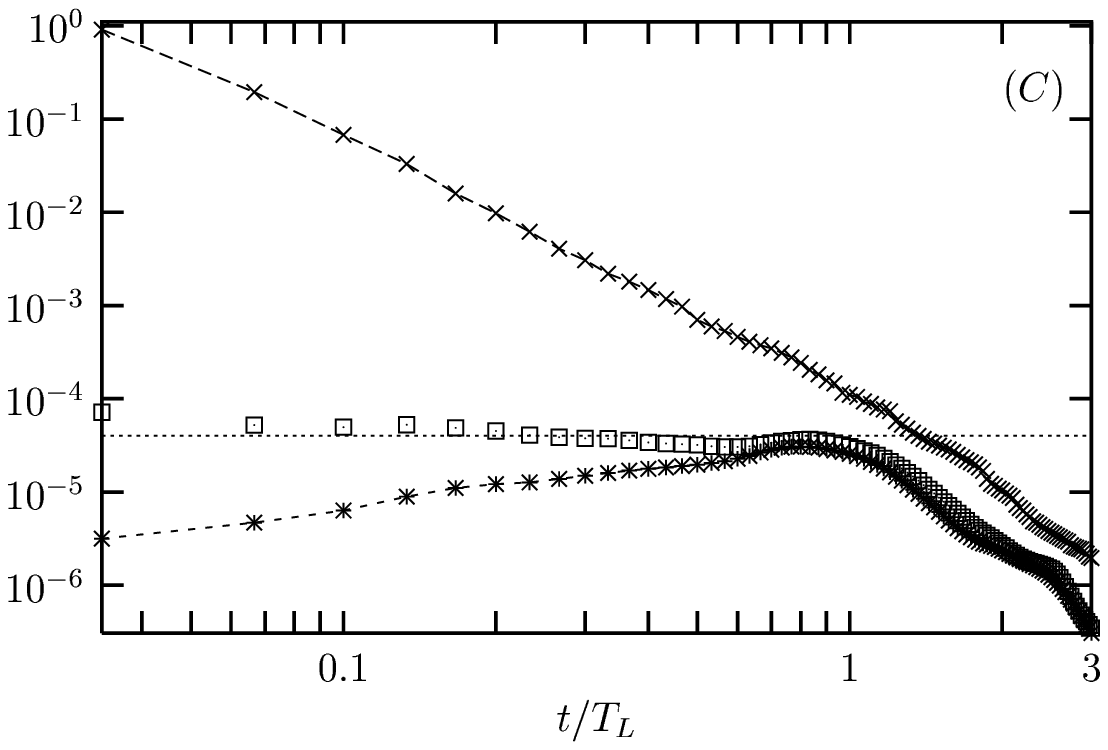}
\caption{Panel (A): time dependence of the decaying second  order
correlation functions ($\times$), together with the time dependence of 
the statistically conserved
quantities $I^{(2)}$, ($\Box$). Equations  (\ref{sabra}) and
(\ref{passive}) have been integrated with a total
number of shells $N=33$. Time in the horizontal-axis is given in units
of the eddy turn over time $T_L$. Panel (B): the same as panel (A)
but for the fourth order correlation function and with $N=25$. 
Panel (C): the same of
panel (B) but for the sixth order correlation function. Here we also present
$I^{(6)}$  when we replace the forced solution
$F_6(n,m,k)$ with its dimensional prediction, ($\ast$) .}
\label{results}
\end{figure}

\end{document}